\documentclass[11pt]{article} 

\usepackage{graphicx,amstext,amssymb}

\begin{document}

\author{S.V. Akkelin$^{1}$, Yu.M. Sinyukov$^{1}$}
\title{HBT search for new states of matter in A+A collisions}
\maketitle

\begin{abstract} A method allowing studies of the hadronic matter
at the early evolution stage in
A+A collisions is developed. It is based on an interferometry
analysis of approximately conserved values such as the averaged
phase-space density (APSD) and the specific entropy of thermal
pions. The plateau found in the APSD behavior vs collision energy
at SPS is associated, apparently, with the deconfinement phase
transition at low SPS energies; a saturation of this quantity at
the RHIC energies indicates the limiting Hagedorn temperature for
hadronic matter. It is shown that if the cubic power of effective
temperature  of pion transverse spectra grows with energy
similarly to the rapidity density (that is roughly consistent with
experimental data), then the interferometry volume is inverse
proportional to the pion APSD that is about a constant because of
limiting Hagedorn temperature. This sheds light on the HBT puzzle.
\end{abstract}

\begin{center}
{\small \textit{$^{1}$ Bogolyubov Institute for Theoretical Physics, Kiev
03143, Metrologichna 14b, Ukraine. \\[0pt]
}}


\end{center}

\section{Introduction}

The main goal of experiments with ultra-relativistic heavy ion
collisions is to study the new forms of strongly interacting matter
which can be created under the extreme conditions. High densities
and temperatures that arise in quasi-macroscopic systems formed
during collision processes can result in the phase transitions from
hadronic gas (HG) to Quark-Gluon Plasma (QGP) or sQGP
\cite{Shuryak1};  the initial very dense pre-thermal stage of the
collisions is, apparently, associated at RHIC energies with a
specific form of matter - Color Glass Condensate (CGC)
\cite{McLerran}. The bulk of hadronic observables ("soft physics")
are related, however, only to the very last period of the matter
evolution, so called thermal or kinetic freeze-out - the end of the
collective expansion of hadronic gas when the system decays. The
evolution of wave function of a single particle, e.g. pion which can
be created, annihilated and scattered  many times in a dense
stochastic (incoherent) surrounding, can be hardly considered. In
fact, the single particle spectra bring information about the state
of matter, e.g., its temperature and flows, at the very end of the
hadron gas evolution, and the HBT interferometry \cite{HBT} is
related directly to the structure of emission function, i.e. the
space-time density of last hadronic collisions. Thus, the
correlation measurements itself cannot be related directly to the
preceding hot and dense stages of the matter evolution in A+A
collisions where a formation of new forms of matter is expected. Our
basic idea  is to use the ''conserved observables'' which are
specific functionals of spectra and correlation functions -
integrals of motion. Then such an observables, being conserved
during the matter evolution (or some period of the evolution) can be
related to the state of matter at the early evolution stage.

A structure of the integrals of motion, besides the trivial ones
associated with the energy, momentum and charges, depends strongly
on a scenario of the matter evolution. The actual numbers of
partonic degrees of freedom released in ultra-relativistic
nuclear-nucleus collisions are fairly big, up to tens of
thousands. As it is commonly supposed such a quasi-macroscopic
system becomes thermal at the early stage of the collision
process: the time of thermalization for the RHIC energies is
estimated to be from $0.6$ fm/c (the phenomenology analysis
\cite{time1}) to 1 fm/c (the ''black hole thermalization''
\cite{time2}) and even more, 3 fm/c (the pure pQCD result
\cite{time3}). Then the system expands nearly isoentropically -
the latter is standard point in the hydrodynamic approach to A+A
collisions and is advocated theoretically for the RHIC energies
\cite{Shuryak1,Shuryak}. As for the SPS energies, the
approximation of perfect hydrodynamics successfully describes the
spectra and correlations while overestimates the elliptic flows
\cite{Heinz}. The latter can be connected with some viscosity
effects at this energies. Keeping this in mind, one can,
nevertheless, consider, at least approximately, the entropy
produced in A+A collisions as integral of motion that carries out
the information as for very initial \textit{thermal} stage of
these processes. The another ''conserved observable'' is found
recently \cite{AkkSin}: it is the pion phase-space density
$\left\langle f\right\rangle $ averaged over momentum (totally or
at fixed rapidity) and configuration space. It is also about a
constant during the stage of chemically frozen expansion. So,
measurements of the entropy and average phase-space density (APSD)
in thermal hadronic systems at the final, freeze-out stage of A+A
collisions makes it possible to look into the previous stages such
as the partonic thermalization and the hadronization (or chemical
freeze-out).

In the paper we express the entropy and APSD of thermal pions
through the observed spectra and interferometry radii irrespective
of unknown form of freeze-out (isothermal) hypersurface and
transverse flows developed based on the approach proposed  in the
Ref. \cite{AkkSin}.\footnote{Unlike many authors (see, e.g.,
\cite{Ferenc,Tomasik}) we utilize almost totally averaged value of
the APSD, $\left\langle f\right\rangle $, instead of
momentum-dependent one $\left\langle f(p)\right\rangle $ because
the extracted value of the later quantity is affected essentially
by assumed form of freeze-out hypersurface as well as a assumed
profile and intensity of flow on it, see Ref. \cite{Tomasik} and
correspondent discussion in Ref. \cite{AkkSin}.} Our aim is to
study the properties of the matter around hadronization stage at
different energies of A+A collisions, from AGS to RHIC, provide
the general analysis of the results and to make conclusions as for
possible new forms of matter formed in these processes.

\section{The entropy and APSD as observables in A+A collisions}

To clarify the problem let us start from the entropy of thermal
pions. As well known the expression for entropy of a gas of
bosons/fermions has the following covariant form
\begin{equation}
S=(2J+1)\int \frac{d\sigma ^{\mu }p_{\mu }d^{3}{p}}{(2\pi )^{3}p^{0}}%
\,[-(2\pi )^{3}f\ln ((2\pi )^{3}f)\pm (1\pm (2\pi )^{3}f)\ln (1\pm
(2\pi )^{3}f)],  \label{ent-def}
\end{equation}
where $\pm $ sign corresponds to bosons/fermions with total spin
$J$ and $\sigma $ is some hypersurface in Minkowski space. The
value depends on the distribution function, or the phase-space
density $f(x,p)$, that should be known to make the corresponding
estimates. It is easy to show, however, that the phase-space
density, e.g. of $\pi ^{-}$, cannot be extracted from the two- and
many- particle spectra even if the system at kinetic freeze-out is
characterized by the locally equilibrated distribution function.
To make it clear let us write the Wigner function (an analogy of
the phase-space density for the quantum systems) for weekly
interacting particles in the mass-shell approximation
\cite{Groot}:
\begin{equation}
f(x,p)=(2\pi )^{-3}\int d^{4}q\delta (q\cdot p)e^{-iqx}\left\langle
a^{+}(p-(1/2)q)a(p+(1/2)q)\right\rangle .  \label{Wigner1}
\end{equation}
Here the brackets $\left\langle ...\right\rangle $ mean the
averaging of the product of creation and annihilation operators
with a density matrix referred to the space-like hypersurfaces
where particles become or are already nearly free. If the
freeze-out is sudden, one uses usually a thermal density matrix at
the freeze-out hypersurface. The invariant single- and double-
(identical) particle spectra have the forms:
\begin{equation}
n(p)=\left\langle a^{+}(p)a(p)\right\rangle
,n(p_{1},p_{2})=n(p_{1})n(p_{2})+\left| \left\langle
a^{+}(p_{1})a(p_{2})\right\rangle \right| ^{2},  \label{spectra1}
\end{equation}
and the correlation function is defined as the following
\begin{equation}
C(p,q)=n(p_{1},p_{2})/n(p_{1})n(p_{2}).  \label{corr-definition}
\end{equation}
It is easy to see from Eq. (\ref{spectra1}) that the complex phase
of the two- operator average $\left\langle
a^{+}(p_{1})a(p_{2})\right\rangle $ cannot be extracted from the
single- and double- particle spectra. It is possible to show that
the same takes place even when many-particle spectra are included
into an analysis. Therefore one cannot reconstruct the
distribution function $f(x,p)$, see Eq. (\ref{Wigner1}), in a
model independent way. However, since $f(x,p)$ is real (note, that
for the locally equilibrated quantum systems the Wigner function
is positive), one can use Eq. (\ref{Wigner1}) to express the
$f^{2}(x,p)$ integrated over space coordinates just through
squared absolute value of the two- operator average, $\left|
\left\langle a^{+}(p-(1/2)q)a(p+(1/2)q)\right\rangle \right|
^{2}$, integrated over $q$. Then, accounting for the direct links
(\ref{spectra1}), (\ref{corr-definition}) of the latter value with
the single particle spectrum (that is just the $f(x,p)$ integrated
over space coordinates) and correlation function, one can get the
phase-space density averaged over some hypersurface $\sigma $,
where all particles  are already free:  $\sigma = \sigma _{out}$,
and over  momentum at fixed particle rapidity, $y=0$,
\begin{equation}
\left\langle f(\sigma ,y)\right\rangle _{y\simeq 0}=\frac{\int
\left( f(x,p)\right) ^{2}p^{\mu }d\sigma _{\mu }d^{2}p_{T}}{\int
f(x,p)p^{\mu }d\sigma _{\mu }d^{2}p_{T}}=\frac{(2\pi )^{-3}\int
p_{0}^{-1}n^{2}(p)(C(p,q)-1)d^{3}qd^{2}p_{T}}{dN/dy},
\label{covariant}
\end{equation}
directly from the experimental data in full accordance with the
pioneer Bertsch idea \cite{Bertsch}. Using the standard
Bertsch-Pratt parametrization the last equality in
(\ref{covariant}) can be re-written through the interferometry
radii, as it is presented in Eq. (\ref{result1}).

On the face of it, the extracted value of the APSD
(\ref{covariant}) does not help to determine the entropy
(\ref{ent-def}), and so, some phenomenological functions that can
reproduce the approximate Gaussian behavior of the correlation
function are usually suggested \cite{Pratt}. However, since we
cannot extract the  phase of the two- operator average, there is
the infinite set of distribution functions compatible with the
observables and, therefore, the entropy calculated will depend on
the class of the functions we choose which. Here we propose the
method to estimate the entropy using just the APSD without
assuming  any concrete expression for the phase-space density.

The method is similar to what was proposed in Ref. \cite{AkkSin}
for analysis of the overpopulation of the phase-space. The idea is
based on the standard approach for spectra formation \cite{Landau}
that supposes the thermal freeze-out in expanding (with 4-velocity
field $u^{\nu}$) locally equilibrated system happens  at some
space-time hypersurface with uniform temperature $T$ and particle
number density $n$ (or chemical potential $\mu$). Then, within
this approximation which is probably appropriate in some
''boost-invariant'' mid-rapidity interval, $\Delta y \lesssim 1$,
the phase-space density of pions,
\begin{equation}
f=f_{l.eq.}(x,p)=(2\pi)^{-3}(\exp(\frac{u_{\nu}(x)p^{\nu}-\mu}{T})-1)^{-1},
\label{f-l.eq.}
\end{equation}
 totally averaged over the
hypersurface of thermal freeze-out, $\sigma =\sigma _{th}$, and
momentum except the longitudinal one (rapidity is fixed, e.g.,
$y=0$) will be the same as the totally averaged phase-space
density in the static homogeneous Bose gas \cite{AkkSin}:
\begin{equation}
(2\pi )^{3}\left\langle f(\sigma ,y)\right\rangle _{y=0}=\frac{\int d^{3}p%
\overline{f}_{eq}^{2}}{\int d^{3}p\overline{f}_{eq}}=\kappa
\frac{2\pi ^{5/2}\int \left( \frac{1}{R_{O}R_{S}R_{L}}\left(
\frac{d^{2}N}{2\pi m_{T}dm_{T}dy}\right) ^{2}\right)
dm_{T}}{dN/dy},  \label{result1}
\end{equation}
where
\begin{equation}
\overline{f}_{eq}\equiv (\exp (\beta (p_{0}-\mu )-1)^{-1}  \label{def1}
\end{equation}
and $\beta $ and $\mu $ coincide with the inverse of the temperature
and chemical potential at the freeze-out hypersurface.\footnote{Note
here that $(2\pi )^{3} f$ is unitless not only in  the natural units
where $\hbar=c=1$ but also in conventional system of units (e.g. SI)
where $(2\pi )^{3} f$ is expressed via the Plank constant: $(2\pi
\hbar)^{3} f=h^{3}f$. Therefore the above value can be interpreted
as distribution of the population numbers in elementary phase-space
cells unlike phase-space density $f$ that carries units $h^{-3}$.}
The $\kappa =1$ \textit{if } one ignores the resonance decays. Here
we neglect interferometry cross-terms since they are usually rather
small in the mid-rapidity region for symmetric heavy ion central
collisions at high energies.

The last expression in Eq. (\ref{result1}) just corresponds to the
last term in  (\ref{covariant}) calculated in the Bertsch-Pratt
parametrization. The important result presented by the first
equality in (\ref{result1}) is based on the properties of
relativistic invariance of the distribution function
(\ref{f-l.eq.}) and its local isotropy in momentum in the rest
frames of each fluid element. Then, using the "boost-invariance"
within homogeneity length, $\Delta y\approx 1$,  the integrals
over $p^{\mu }d\sigma _{\mu }d^{2}p_{T}$ of \textit{diverse}
functions $F_{(i)}$ of the locally-equilibrium distribution,
$F_{(i)}(f_{l.eq.})$, contain the common
factor, ''effective volume'' $V_{eff}=\int \frac{d\sigma _{\mu }}{d\eta }u^{\mu }$ ($%
\eta $ is rapidity of fluid), that completely absorbs the flows
$u^{\mu }(x)$ and form of hypersurface $\sigma (x)$ in
mid-rapidity. For instance, if $F_{(i)}(f_{l.eq.})=f_{l.eq.}$,
then
\begin{equation}
\frac{dN}{dy}=\int p^{\mu }d\sigma _{\mu }d^{2}p_{T}f_{l.eq.}
=V_{eff}\int d^{3}p\frac{\overline{f}_{eq}}{(2\pi)^{3}}=
n_{th}V_{eff} \label{V-eff}
\end{equation}
where $n_{th}$ is thermal density of equilibrium ideal Bose gas,
and similar takes place for $F_{(j)}(f_{l.eq.})=f_{l.eq.}^2$ in
(\ref{covariant}). Thus, the effective volume $V_{eff}$ is
cancelled in the corresponding ratios. This factorization property
has been found first for Eq. (\ref{V-eff}) in Refs.
\cite{Nukleonika,Cleymans1}, multiple used for an analysis of
particle number ratios (see, e.g., Ref. \cite{A-B-S}) and recently
generalized for a study of the APSD in Ref. \cite{AkkSin}.

In this work we apply the approach to an analysis of the thermal
pion entropy per particle, or specific entropy of thermal pions.
Using the same approximation of the uniform freeze-out temperature
and density and Eq. (\ref {ent-def}) with some local equilibrium
distribution we get the following expression for specific entropy
in mid-rapidity:
\begin{equation}
\frac{dS/dy}{dN/dy}=\frac{\int d^{3}{p}\,[-\overline{f}_{eq}\ln \overline{f}%
_{eq}+(1+\overline{f}_{eq})\ln (1+\overline{f}_{eq})]}{\int d^{3}p\overline{f%
}_{eq}}.  \label{en-to-n}
\end{equation}
In the above ratio due to the factorization property the effective
volume is cancelled and the final expression depends only on the
two parameters: the temperature and chemical potential at
freeze-out. The temperature can be obtained from the fit of the
transverse spectra for \textit{different} particle species and we
will use the value $T=120$ MeV as a typical "average value" for
SPS and RHIC experiments.

Another parameter, the chemical potential, cannot be extracted
from the spectra, its value could be fairly high even for the
thermal pions because of chemical freeze-out and this parameter is
crucial for an estimate of the entropy. We extract the chemical
potential from an analysis of the APSD  following to
(\ref{result1}).  The factor $\kappa $  is accounting for a
contribution of the short-lived resonances to the spectra and
interferometry radii and absorbs also the effect of suppression of
the correlation function due to the long-lived resonances
\cite{AkkSin}. Because of the chemical freeze-out a big part of
pions, about a half, are produced by the short-lived resonances
after thermal freeze-out. It leads to an increase of the APSD
despite the maximal particle and entropy densities of pions at
post-hydrodynamic stage of the evolution is reached  at the end of
the hydrodynamic expansion - at thermal freeze-out as  discussed
in  detail in Ref. \cite{AkkSin}.  To estimate the thermal
characteristics and "conserved observables" at the final stage of
hydrodynamic evolution by means of Eqs. (\ref{result1}),
(\ref{en-to-n}) one needs to eliminate non-thermal contributions
to the pion spectra and correlation functions from resonance
decays at post freeze-out stage.  To do this we use the results of
Ref. \cite{AkkSin} where a study of the corresponding
contributions within hydrodynamic approach gives the values of
parameter $\kappa$ to be $\kappa =0.65$ for SPS and $\kappa =0.7$
for RHIC, if half of pions is produced by the resonances at post
freeze-out stage. Then,  from Eqs. (\ref{result1}), (\ref{def1})
one can extract the pion chemical potential at thermal freeze-out.
This makes it possible to estimate the average phase-space
density, the specific entropy of thermal pions and other thermal
parameters of the system at the end of the hydrodynamic expansion
as  explained above.
\section{The analysis of experimental data and the results}
To evaluate the APSD of negative  pions by means of Eq.
(\ref{result1}) we utilize the yields, transverse momentum spectra
and interferometry radii of $\pi ^{-}$ at mid-rapidity measured in
central heavy ions collisions by the  E895 and E802 Collaborations
for AGS energies \cite{E895,E802}, NA49 Collaboration for SPS CERN
energies \cite{NA49,NA49-report,NA49-1}, STAR and PHENIX
Collaborations for RHIC BNL energies \cite{STAR,PHENIX,PHENIX1}.

 To calculate the APSD we have to integrate over the whole
$p_T$ region. Always, if possible, we use the pion yields and
values of radii in each measured $p_T$ bin instead of analytical
parameterizations for the transverse spectra and interferometry
radii. To interpolate between successive data points we use an
polynomial functions, the degree of the polynomial curves is
chosen to be $2$ or $3$. Outside the whole measured $p_T$ region
we make analytical extrapolations with parameters that catch the
main tendencies of the data points in the measured $p_{T}$ region.
We utilize  the analytical parameterizations that are typically
used in fitting procedures and  are motivated by hydro model
calculations. Namely, for the transverse spectra at high $p_T$ we
assume the exponential parameterizations with slopes that
correspond to average ones in the measured $p_T$ region (that is,
actually, the average slope in the measured $p_T\lesssim 1$ GeV
points). For small unmeasured $p_T$ our exponential extrapolations
are taken with the slopes which are defined from a requirement of
coincidence of the resulting total $\pi ^{-}$ yields at
mid-rapidity with the values presented by experimental
Collaborations. To extrapolate interferometry radii behind the
measured $p_{T}$ bins, we utilize the widely used simple
analytical parameterizations for interferometry radii,
$a/(b+c*m_{T})^{d}$, with numerical parameters which are taken, if
possible, from the fit given by the experimental Collaboration, or
we determine them ourselves if they are not presented. Also, we
use the approximation $R_{O}=R_{S}$ for the outward and sideward
interferometry radii below the minimal measured $p_{T}$ momentum,
because this equality should take place for $p_{T}=0$ due to
evident (and well known) symmetry reasons. We neglect
interferometry cross-terms because they are rather small at
mid-rapidity (see, e.g., \cite{NA49-1})), and, for example, at SPS
energies $\sqrt{R_{O}^{2}R_{L}^{2}-R_{O,L}^{2}}$ is equal to
$R_{O}R_{L}$ with a few percents accuracy.  Since the
interferometry radii are measured by PHENIX Collaboration for
$0-30$ $\%$ centrality events at $\sqrt{s_{NN}}=200$ GeV
\cite{PHENIX1}, we increase the interferometry volume measured by
PHENIX Collaboration at this c.m. energy  by a factor of  $1.215$
to get the interferometry volume corresponding to the most central
$0-5$ $\%$ centrality bin in accordance with $N_{part}$ dependence
of the Bertsch-Pratt radius parameters found in Ref.
\cite{PHENIX1}.

The results for the APSD at mid-rapidity for \textit{all} negative
pions at the AGS, SPS, RHIC energies in logarithmic c.m. energy
scale are presented in Fig. 1. Note that for SPS energy domain we
use the results of NA49 Collaboration since it presents
interferometry data as well as the data on transverse momentum
spectra which are necessary to get the APSD values, while the
CERES Collaboration demonstrates the interferometry radii only.

The non-monotonic structure of the APSD behavior seen in Fig. 1 in
the AGS-SPS energy domain could be an indicator of new physical
phenomena as it is discussed in next Section, so a statistical
significance of this structure is important and we demonstrate it
in Fig. 2 at AGS, SPS energies in $\sqrt{s_{NN}}$  scale. The
comment is the following. Because of fast (exponential) decrease
in of transverse momentum spectra with $m_{T}$ the uncertainties
of the spectra and HBT radii in the region of high $p_{T}$ has
only a little influence on the calculated APSD. The latter are
mostly affected by the values of the HBT radii at low $p_{T}$. The
most important are systematic (from a choice of analytic $p_{T}$
parameterizations) uncertainties in this region behind the lower
measured $p_{T}$ bin, and experimental errors in the HBT data.
Taking into account that the experimental errors in the HBT radii
do not exceed typically $0.5$ fm (see, e.g., \cite{NA49-1}), and
estimating the resulting systematic uncertainties from our choice
of $p_{T}$ parameterizations as $0.5$ fm, we get the uncertainty
of $1$ fm in each HBT radius. For a rough estimate of relative
errors in the APSD values let us assume that average value of HBT
radius in the region of interest is approximately $5$ fm. Then,
supposing that all uncertainties are independent and accidental,
we can estimate that those errors are approximately $30$ percents.
As one can see from Fig. 2, the APSD in AGS energy domain can be
approximated with quite good accuracy by linear in $\sqrt{s_{NN}}$
function, $a \sqrt{s_{NN}} + b$, where $a=0.03768$ GeV$^{-1/2}$
and $b=0.01345$. Then the extrapolation of this tendency   to SPS
energy domain results in rather high values for APSD, for example,
at $\sqrt{s_{NN}}=12.3$ GeV we get $(2\pi )^{3}\left\langle
f\right\rangle = 0.477$ that is far from the estimated error bars,
$(2\pi )^{3}\left\langle f\right\rangle = 0.203 \pm 0.06$. While
the final conclusion about statistical significance of the
observed tendencies needs in more detail studies both theoretical
and experimental, our estimate from the above analysis is that it
is very likely that experimental data really indicate
non-monotonic behavior of the APSD as a function of collision
energy in the AGS-SPS energy domain.

In Fig. 3 the APSD of \textit{thermal} negative pions at the SPS
and RHIC energies are presented. Here and below we demonstrate the
values for the thermal negative pions at the RHIC energies which
are mean values of STAR and PHENIX data. Note, that in Fig. 3 we
present also the APSD of negative thermal pions at chemical
freeze-out in assumption of chemical equilibrium, ($\mu_{\pi}=0$).
We use the temperatures of chemical freeze-out for different
colliding energies  from Refs. \cite{chemic}, where they were
found from analysis of particle number ratios.  We do not
calculate the APSD of \textit{thermal} pions at the AGS energies
because estimates of resonance contributions to the pion spectra
have been done in \cite{AkkSin} for the SPS and RHIC energies
only.

The  APSD of negative thermal pions are used then to extract the
chemical potentials $\mu$ of them at thermal freeze-out with
$T_{th}=120$ MeV at different SPS and RHIC energies, and after
that to calculate the specific entropies, $s= \frac{dS/dy}{dN/dy}$
(\ref{en-to-n}), the entropies, $\frac{dS}{dy}=s*\frac{dN}{dy}$,
and the densities $n_{th}$, see Eq. (\ref{V-eff}), of negative
thermal pions. The values founded are presented in Figs. 4, 6, 7.
Also we demonstrate in Fig. 5   the interferometry volume
$V_{int}=(2\pi)^{3/2}R_{O}R_{S}R_{L}$ calculated at small
$p_{T}\simeq 0.06 - 0.07$ GeV as the function of the rapidity
densities, $dN^{\pi^{-}}/dy$, of all negative pions  at
mid-rapidity in central nucleus-nucleus collisions. For the RHIC
energies we use in Fig. 5 the interferometry radii measured by
STAR Collaboration because the Collaboration presents the
interferometry measurements in lower $p_{T}$ bins as compare to
PHENIX Collaboration. If there are no experimental data in
selected $p_T$ bin, $p_{T}\simeq 0.06 - 0.07$ GeV, we calculated
the correspondent values using the analytical parameterizations of
the interferometry radii as explained above. The obtained values,
$V_{int}$, are used then to evaluate the ratios
$(dN^{\pi^{-}}/dy)/V_{int}$ that are demonstrated and compared
with the thermal densities $n_{th}$ (\ref{V-eff}) in Fig. 6.

In Fig. 8 we present, in addition to Fig. 7 where the entropies
$dS/dy$ of negative thermal pions are demonstrated, the rapidity
densities, $dN^{\pi^{-}}/dy$, of all negative pions at
mid-rapidity in central nucleus-nucleus collisions for the AGS,
SPS and RHIC energies. The experimental values for the pion
rapidity densities are taken from Refs.
\cite{E895,E802,NA49,NA49-report,STAR,PHENIX}. We used the
rapidity densities of pions instead of those for \textit{all }
charged particles, since pions are the new produced particles
which are not contained initially in colliding nuclei, and,
therefore, more directly represent a mechanism of the particle
production in A+A collisions. It is especially important for
collision processes with relatively low multiplicities at the AGS
energies, where a large fraction of registered charged particles
are protons which were not produced in collision processes being
initially in colliding nuclei. The lines in Fig. 8 represent the
logarithmic law of energy dependence for negative pion
multiplicities: $a\log_{10} (\sqrt{s_{NN}}/b)$, where $a=160
(230)$, $b = 1.91$ GeV (3 GeV) for solid (dashed) lines
respectively. Note, that we use the STAR point
$dN^{\pi^{-}}/dy=249$ for $\sqrt{s_{NN}}=130$ GeV because this
value - the result of a use of the Bose-Einstein distribution as a
fit function, is closer then another STAR value based on a
Boltzmann-like fit (see detail in \cite{STAR}) to the value
$dN^{\pi^{-}}/dy=270\pm 3.5$ reported by PHENIX Collaboration and
is also closest to  the result $dN^{\pi^{-}}/dy=287\pm 20$ that is
deduced by STAR Collaboration from their measurements of negative
hadrons, antiprotons and negative kaons spectra in Ref.
\cite{STAR-1}. Note that the latter value (it is not presented in
Fig. 8) is even closer  to the fitting line, which is showed in
Fig. 8 as the solid line, than demonstrated experimental value.

\section{Discussion and interpretation of the results}
Let us start from an analysis of the $\sqrt{s_{NN}}$ dependance of
the averaged phase-space density per unit of rapidity (APSD).
Figure 1 is based on Eq. (\ref{covariant})  or the last term in
Eq. (\ref{result1}) with $\kappa = 1$ and demonstrates a behavior
of the "raw" APSDs (which are, actually, the "asymptotic" APSDs
related to the times when pions are detected) accounting for all
negative pions, $\pi ^-$, thermal and from resonance decays at
post thermal stage. One can see that the APSD grows significantly
with energy at the AGS energies, then has the plateau starting
from the lowest SPS energy, 20 AGeV, till 80 AGeV and then begins
to grow again, apparently very slowly at RHIC as one can conclude
from the non quite compatible experimental data of the STAR and
PHENIX Collaborations.

Unlike a fast decrease of  particle $n(x)$ and phase-space
$f(x,p)$ local densities, the totally averaged phase-space
densities  of thermal pions are conserved during a chemically
frozen evolution \cite{AkkSin} Roughly, it is proportional to the
total ("raw") APSD at the SPS and RHIC energies, see Fig. 3. If
the same properties take place at the AGS energies too, then one
can easily interpret the behavior of the averaged phase-space
density in Fig. 1. When the energy of collisions at AGS grows, the
initial hadronic density and phase-space density increase; since
the pion APSD is conserved, its observed value also grows.  Its
rise stops at low SPS energies, this means that the initial
density of pions also stops increasing. The simplest explanation
is: an excess of the initial energy begins to be transformed into
new, non pionic (hadronic) degrees of freedom, possible to quarks
and gluons. The pure hadronic stage appears later when the
densities became smaller than the initial ones, therefore the
initial APSD of pions depends then not on the initial energy
density but on the density determined by the hadronization
temperature $T_c$. Figure 3 demonstrates that the APSD of the
\textit{thermal } pions  grows indeed very slowly starting from
high SPS energies that can reflect the fact of saturation of the
temperature of the phase transition at the RHIC energies. The APSD
at the thermal freeze-out is slightly higher than at chemical one.
It is because the conservation of the APSD should take place at
perfectly chemically frozen hadronic evolution; there is, however,
a residual effect of increase of pion number  because of an excess
of resonance decays into the expanding gas over back processes of
the recombination. This difference does not contradict to typical
estimates that roughly $2/3$ of pions at hadronization stage are
"hidden" in resonances
\cite{Braun-Munzinger,Braun-Munzinger1,A-B-S} and about half of
pions are already thermal to the end of the hydro evolution. The
pion APSDs at the chemical freeze-out are calculated using the
thermal parameters of that stage that were found from an analysis
of particle number ratios at SPS and RHIC in Refs. \cite{chemic}.
A dependance of the chemical potential $\mu$ of thermal pions at
thermal freeze-out  on $\sqrt{s_{NN}}$ is demonstrated in Fig. 4.
One can see that the chemical potential of thermal pions at
freeze-out is saturated somewhere  between 50 and 60 MeV.

 Figures 5 and 6 are related to the behavior of the interferometry
volume $V_{int}$ on multiplicities, $dN^{\pi^{-}}/dy$, in central
collisions at different energies of AGS, SPS, and RHIC.  The
interferometry volume has a tendency to grow over a broad energy
range, as one can see from Fig. 5.  However for {\it central}
Pb+Pb (Au+Au) collisions at different energies, the corresponding
increase is much slow than the proportionality law between
$V_{int}(\sqrt s)$ and $dN^{\pi^{-}}/dy(\sqrt s)$. Also, there is
the statistically reliable violation of a monotonic behavior of
$V_{int}$ \cite{ceres} manifested in the evident decrease of
$V_{int}$ in AGS energy interval. The phenomenon has been
associated with the supposed constancy of the (kinetic) freeze-out
value of pion mean free path while the transition from the nucleon
to pion dominated matter happens within that energy range
\cite{ceres}. Note that the relatively steep rise of the pion
APSDs (see Figs. 1 and 2) at a moderate increase of pionic
multiplicities at the AGS energies (see Fig. 8) is caused by the
discussed decrease of $V_{int}$ with beam energy. Of course, the
conservation of the APSD in absence of the deconfinement phase
transition leads in any case to some rise of its value with
initial energy density.
 As  follows from Fig. 6, the $dN^{\pi^{-}}/dy$  grows with energy
significantly faster than $V_{int}$. This fact is the main
component of the HBT puzzle \cite{puzzle,puzzle1}. To understand
it qualitatively, let us very roughly estimate the APSD
(\ref{result1}) supposing that the transverse spectra have mainly
exponential behavior vs transverse mass $m_{T}$, $ \propto
\exp{(-m_{T}/T_{eff})}$, where effective temperature $T_{eff}$
depends on the thermodynamic temperature at the hypersurface of
thermal freeze-out $\sigma$ and flows at $\sigma$. Then, assuming
that integral $I$ over dimensionless variable $m_{T}/T_{eff}$
depends on energies of collisions fairly smoothly, one can write
\begin{equation}\label{APSD-simple}
V_{int}(\sqrt{s}) \simeq I\frac{dN/dy}{\langle f\rangle
T_{eff}^{3}}
\end{equation}
where the interferometry volume is taken here at the smallest
$m_{T}$. Thus, a proportionality between $V_{int}$ and the
particle numbers $dN/dy$ is destroyed by a factor $\langle
f\rangle T_{eff}^{3}$. So, if the APSD and $V_{int}$ only slightly
grow with energy, mostly an increase of $T_{eff}^{3}$ could
compensate a growth of $dN/dy$ in Eq. (\ref{APSD-simple}). One can
see that it is the case: for example, the ratio of cube of
effective temperatures of negative pions  at $\sqrt{s_{NN}}=200$
GeV (RHIC) to one at 40 AGeV (CERN SPS) gives approximately $2$,
while the ratio of corresponding  mid-rapidity densities is
approximately equal to 3 \cite{NA49,STAR,PHENIX}.  It could be
only in the case of an increase of the pion flows in A+A
collisions with energy. If the intensity of flows grows, it leads
to a reduction of homogeneity lengths and the corresponding
interferometry radii \cite{AkkSin1}. This effect almost
"compensates" a growth of final geometrical sizes of the system
with energy in observed interferometry volumes. Indeed, as one can
see from Fig. 6 the freeze-out densities for pions, $n_{th}$,
become noticeably smaller than formally defined HBT densities,
$\frac{dN/dy}{V_{int}}$, starting from the high SPS energies. In
other words, the interferometry volume at those energies becomes
to be significantly smaller than the effective one occupied by the
system and defined by Eq. (\ref{V-eff}).

The result (\ref{APSD-simple}) brings about some more details. If at
any \textit{fixed} energy $\sqrt{s_{NN}}$ we look at the evolution
in time of $V_{int}$, we found that it can be nearly constant since
the values $dN/dy$, APSD $\langle f \rangle $ and effective
temperature $T_{eff}$ in r.h.s. of Eq. (\ref{APSD-simple}) are
approximately conserved for the thermal pions during the chemically
frozen hydro-evolution \cite{AkkSin}. As the result, the "HBT
microscope"  at diverse  energies "measures" the radii that are
similar to the sizes of colliding nuclei. It provides an explanation
to the phenomenological observations \cite{NA49-1,Appel} that in
central Pb+Pb and Au+Au collisions the interferometry volumes grow
rather slowly with energy, and only due to the longitudinal
interferometry radius grows (transverse sizes are equal),  while, at
the same energy, the $V_{int}$ depends strongly on the sizes of
colliding nuclei and on the impact parameters in non-central
collisions.

 Our other observations are based on one more conserved
value, the entropy. In principle, the entropy $S$ of the  pion
component alone can be changed even in the perfect fluid as well
as the pion numbers $N$ (see discussion above). One can expect,
however, that such deviations will be small for the ratios
$s=\frac{dS/dy}{dN/dy}$ or for the specific entropy. Then using
the result (\ref{en-to-n}) for the latter value at the thermal
freeze-out and the chemical potential extracted from an APSD
analysis, one can determine the specific entropy of pions. The
corresponding estimates give the values of $s$ to be approximately
equal to 4 at SPS except for the top SPS energy, where $s\simeq
3.79$. At RHIC energies for $\sqrt{s_{NN}}=130$ GeV, the
corresponding averaged value is equal to $3.66$ (one can conclude
from Fig. 1 that such a low value is, probably, artefact and
result of relatively high discrepancy between the STAR and PHENIX
data), and for $\sqrt{s_{NN}}=200$ GeV the specific entropy of
negative pions is equal to $3.82$.

 The total entropy
of thermal negative pions per unit of rapidity, that is  $s$
multiplied by the rapidity densities of \textit{thermal} pions,
$(dN^{\pi^{-}}/dy)/2$, is presented in Fig. 7. One can see that the
entropy starts to grow faster at the SPS energies but at the RHIC
energies this tendency is canceled.  To understand better the
situation let us look at the "raw" data in Fig. 8 representing a
behavior of the negative pion rapidity density in the AGS, SPS and
RHIC experiments. From this picture, that makes  the tendencies
which are seen in Fig. 7 more evident, one can conclude that at the
SPS energies there is, indeed, an anomalously large slope of an
increase of the pion entropy (and the number of pions) with energy.
The observed multiplicities of negative pions at SPS at 80 AGeV (158
AGeV) are by factors of 1.09 (1.143) larger than ones extrapolated
in accordance with the tendency (dashed line) observed at the AGS
energies and, apparently, at the RHIC energies where relatively low
pion multiplicities could be remnants of CGC formation
\cite{McLerran}. Taking into account that chemically equilibrated
thermal model correctly describes particle number ratios, including
pions, at the AGS and RHIC energies \cite{Braun-Munzinger1} and that
at the top SPS energy there is a problem of a "pion deficit"
\cite{A-B-S}\footnote{The problem of pion deficit in mixed (full
$4\pi $ geometry - midrapidity) particle number ratios in central
158 AGeV Pb+Pb collisions was pointed out earlier in Ref. \cite{Gor}
were preliminary data on hadron multiplicities  were analyzed within
hadron gas models.} we can suppose that the observed "excess" of
pions is caused by a mechanism shifting the pion production at the
SPS energies from the equilibrium. This mechanism could be also in
some degree responsible for the reduction of K/$\pi$ ratios as
compare to ones in chemical equilibrium model \cite{Cleymans} - the
effect was observed by NA49 Collaboration ("horn" puzzle)
\cite{NA49-report}. Some increase of the pion APSD at SPS 158 AGeV,
see Figs. 1, 2, 3 could be also caused by the "extra pion"
production at high SPS energies.

What could be the reason of the "extra pion" production out of
equilibrium at SPS energies? The intriguing possibility is that
such an effect is the manifestation of the QCD critical end point
(CEP) which is the terminating point of the first order phase
transition line (about physics of the CEP and its location at QCD
phase diagram see, e.g., \cite{Stephanov1}). Indeed, since the CEP
acts as an attractor of the isentropic trajectories of the
thermodynamic system evolution \cite{Asakawa}, the critical domain
can influence the particle spectra for some range of collision
energies, e.g., this could be responsible for the "step" behavior
of the kaon effective temperature, discovered by NA49
Collaboration at the SPS energies \cite{NA49-report}. Then
nonequilibrium features, which accompany the phase transition in
the expanding systems, e.g., a rise of the bulk viscosity in the
mixed phase due to a finite relaxation time and variation of sound
velocity in the transition region \cite{Gyulassy}, could lead to a
dissipation of kinetic energy and to entropy production. Another
effect, that can result in the "extra pion" production in A+A
collisions near the CEP, is a significant reduction (that is
maximal in the vicinity of the CEP at the QCD phase diagram) of
$\sigma $ meson mass from its vacuum value, and mass shift of
other resonances ($\rho$, etc.) due to sigma exchanges
\cite{Stephanov2,Shuryak2}. As a result these species are rather
numerous around the CEP. As $m_{\sigma}<2m_{\pi}$ in a vicinity of
the CEP \cite{Stephanov2,Shuryak2}, such sigma mesons cannot decay
into $\pi\pi$ in this state and can do it only after $\sigma$
meson masses are increased that happens when the density in A+A
collisions are reduced by the system expansion. One can speculate
that when it happens the rate of inelastic collisions can be not
high enough to push the new produced pions into chemical
equilibrium, while the elastic collisions can still thermalize
them. That is why there could be no peculiarities in the pion
transverse momentum spectra in low $p_{T}$ region, as predicted in
Ref. \cite{Stephanov2}, while an enhancement of pion yields can be
considered as a possible manifestation of the CEP in A+A
collisions.\footnote{The "extra pion" production was considered
earlier as the signature of the chiral phase transition in Ref.
\cite{Koch}.} Since the "pion excess" is maximal at the highest
SPS energy 158 AGeV, it could mean, unlike present expectations
\cite{Stephanov1}, that the CEP is situated in the QCD phase
diagram closer to the chemical freeze-out point at highest SPS
energy than at the lowest SPS one. This possibility is argued for
recently in Refs. \cite{Antoniou} and \cite{Gavai}. It is
noteworthy that because of an inhomogeneity of the baryonic
chemical potential and temperature in rapidity at the chemical
freeze-out hypersurface (see, e.g., an analysis that has been done
in Ref. \cite{Sollfrank}), the condition for the thermodynamic
trajectory of system evolution to pass around the CEP could be
realized just in mid-rapidity region. Then the enhancement of pion
production can be observed from diverse particle number ratios in
the unit central rapidity interval rather than from $4\pi$
 abundances.

At the RHIC energies both mechanisms, bulk viscosity and sigma
mass reduction resulting in intensive entropy and pion
multiplicities rise, can be inefficient  since at small net baryon
density, which are typical for that energies, the crossover far
from critical domain around CEP might, apparently, happen (see,
e.g., \cite{crossover}).

A few remarks are in order here. Presently, one of the
interpretation of a larger rate of an increase of pion production
in A+A collisions at the SPS energies as compared to AGS energies
and properly normalized $p+p(\overline{p}$) collisions is based on
the statistical model of the early stage (SMES) \cite{Gorenstein1}
(see also \cite{Gorenstein2}). In this model the kink-like change
(and horn-like structure of $K^{+}/\pi^{+}$ ratios) is a direct
consequence of an onset of the deconfinement and liberation of
massless quark-gluon degrees of freedom in the Lorentz-contracted
fireball at the initial stage of A+A collisions. However it seems
that both basic assumptions, namely,  massless quark-gluon plasma
and the Lorentz-contracted fireball, are not supported  by further
studies. The effective masses of quarks and gluons in the QGP are
temperature-dependent and grow with temperature (see, e.g., Ref.
\cite{Blaizot} where the validity of \textit{quasiparticle}
picture of the quark-gluon plasma at  high temperatures is
advocated) that turns down the possibility to treat the QGP as
weak coupling \textit{massless} quark-gluon system even for $T
\gtrsim 3T_{c}$. It is also clearly seen from the lattice QCD
results \cite{Karsch} where the pressure and energy density are
both below the Stefan-Boltzmann limit even at very high
temperatures which will be hardly reached in heavy ion collisions.
Another crucial assumption in the SMES is a formation
 of the longitudinally
 Lorentz-contacted  fireball in rest, that is actually the Landau-type initial conditions for
hydrodynamic expansion. It was demonstrated \cite{Gorenstein3},
however,  that the Landau-type initial conditions are unable to
reproduce effective temperatures together with other data
(multiplicities and rapidity distributions) at the SPS energies,
and that these quantities can be described altogether only when
one uses large initial volume with an appropriate velocity
distribution (see also \cite{Mohanty}).
\section{Conclusions}
The method allowing  analysis of an early stage of the hadronic
matter evolution in A+A collisions is developed. It is based on
studies of approximate integrals of motions for the evolution of
hadronic systems, such as the totally averaged phase-space density
(APSD) and the  specific entropy of thermal pions. We express
these quantities through experimental data on the spectra and
Bose-Einstein correlations in a way that does not depend on the
freeze-out hypersurface and collective flows developed. Our
estimates of the APSD at hadronization stage are close to the
corresponding ones that we calculate based on the results of
analysis of particle number ratios. A behavior of the pion APSD vs
collision energy has a plateau at low SPS energies that indicates,
apparently, the transformation of initial energy to non-hadronic
forms of matter at SPS; a saturation of that quantity at the RHIC
energies can be treated as an existence of the limiting Hagedorn
temperature of hadronic matter, or maximal temperature of
deconfinement $T_c$. It is noteworthy that observation as for
compatible values of the momentum-dependent APSD $\left\langle
f(p)\right\rangle $ for collisions of different nuclei at top SPS
and top AGS energies has been interpreted in Ref. \cite{Ferenc} as
the universal properties of kinetic freeze-out in heavy ion
collisions. Our study shows, however, that the pion APSD is
approximately conserved value and so has no direct link to the
freeze-out criterion and final thermodynamic parameters, being
connected rather to the initial phase-space density of hadronic
matter in A+A collisions \cite{AkkSin}.

A behavior of the  entropy of thermal  pions and measured pion
multiplicities in central rapidity region vs  energy demonstrates
an anomalously high slope of an increase of the pion
entropy/multiplicities at SPS energies compared to what takes
place at the AGS  and RHIC energies. This additional growth could
be, probably, a manifestation of the QCD critical end point. The
observed phenomenon can be caused by the dissipative effects that
usually accompany phase transitions, such as an increase of the
bulk viscosity, and also by peculiarities of pionic decays of
$\sigma$ mesons and other resonances with masses that are reduced,
as compare to its vacuum values, in vicinity of the QCD CEP. At
the RHIC energies there is no anomalous rise of pion
entropy/multiplicities, apparently, because the crossover
transition takes place far from the CEP and no additional degrees
of freedom appear at that scale of energies: quarks and gluons
were liberated at previous energy scale.

    We also analyze a behavior of the interferometry radii with energy
in a context of the HBT puzzle. We show that if cubic power  of
the effective temperature of pion spectra grows with energy
similar to the rapidity density then the interferometry volume is
inversely proportional to the pion APSD. The behavior of the
latter with collision energy is nearly constant starting from the
high SPS energies  because of the limiting Hagedorn temperature,
$T_{c}$, for hadronic matter. Roughly, the similarity between rise
of effective temperature cubed and pion rapidity density takes
place within the wide interval: from the lowest SPS energy to the
highest RHIC energy. Therefore interferometry volume in Pb+Pb and
Au+Au cental collisions is nearly constant, more precisely, it
grows much slower (mostly due to an increase of longitudinal
radius associated with total lifetime of the system) than rapidity
density. At the same time, at each fixed energy the pion APSD,
rapidity density and effective temperature of pion spectra are
approximately conserved during the evolution \cite{AkkSin},
therefore the interferometry volume, that is the function of above
values, is nearly the same as at the initial moment of hadronic
evolution, if it were measured. It explains  the experimental
observations \cite{NA49-1,Appel} that the interferometry volumes
are changed only a little with energy for central collisions of
the same nuclei and, at the same energy, they are proportional to
the initial system extension in non-central collisions and central
collisions of nuclei with different atomic numbers. The further
experimental analysis of a such type of the correlations between
the interferometry volumes, initial system sizes, multiplicities
and slopes of the transverse spectra are still needed to clarify
the picture.

Summarizing this work, we point out  that the available
interferometry data not only do not contradict  possible dramatic
transformations of the matter in A+A collisions, as it is usually
concluded \cite{puzzle1,puzzle3}, but being analyzed properly give
deep insight into the physics of the phase transitions. The
precise localization of the collision energy region where the
transition to QGP at finite net baryonic densities happens is very
important for understanding of the physics of deconfinement
transition. At the moment no single experiment has collected data
that by themselves show non-monotonic behavior of physical
observables as a function of collision energy in the AGS-SPS
energy domain. This can throw doubt upon an experimental
significance of the observed APSD behavior which leads to the
treatment of that as the result of a phase transition and critical
end point. In view of this the Compressed Baryonic Matter (CBM)
project \cite{CBM} at the future Facility for Antiproton and Ion
Research (FAIR) in Darmstadt is particularly important because it
makes possible the systematic studies of heavy ion collisions at
beam energy range between $10$ and $40$ AGeV.  Our results as for
the localization of the QCD CEP  indicate the importance of the
experiments with relatively light nuclei at the top SPS energy and
future RHIC energy scans in which the corresponding part of the
($T - \mu_{B}$) plane can be reached. Also, a testing of our
prediction of the APSD  suturation at the top RHIC and LHC
energies because of the limiting Hagedorn temperature, associated
with the deconfinement temperature at zero net baryonic densities,
is of a great importance.

\section*{Acknowledgments}

We are grateful  to M. Gyulassy and L. McLerran for their interest
in this work and fruitful discussions. The research described in
this publication was made possible in part by Award No.
UKP1-2613-KV-04 of the U.S. Civilian Research $\&$ Development
Foundation for the Independent States of the Former Soviet Union
(CRDF). Research carried out within the scope of the ERG (GDRE):
Heavy ions at ultrarelativistic energies -– a European Research
Group comprising IN2P3/CNRS, Ecole des Mines de Nantes, Universite
de Nantes, Warsaw University of Technology, JINR Dubna, ITEP
Moscow and Bogolyubov Institute for Theoretical Physics NAS of
Ukraine. The work was also supported by NATO Collaborative Linkage
Grant No. PST.CLG.980086 and Fundamental Researches State Fund of
Ukraine, Agreement No. F7/209-2004.

\newpage

\begin{figure}[h]
\centering
\includegraphics[scale=0.5]{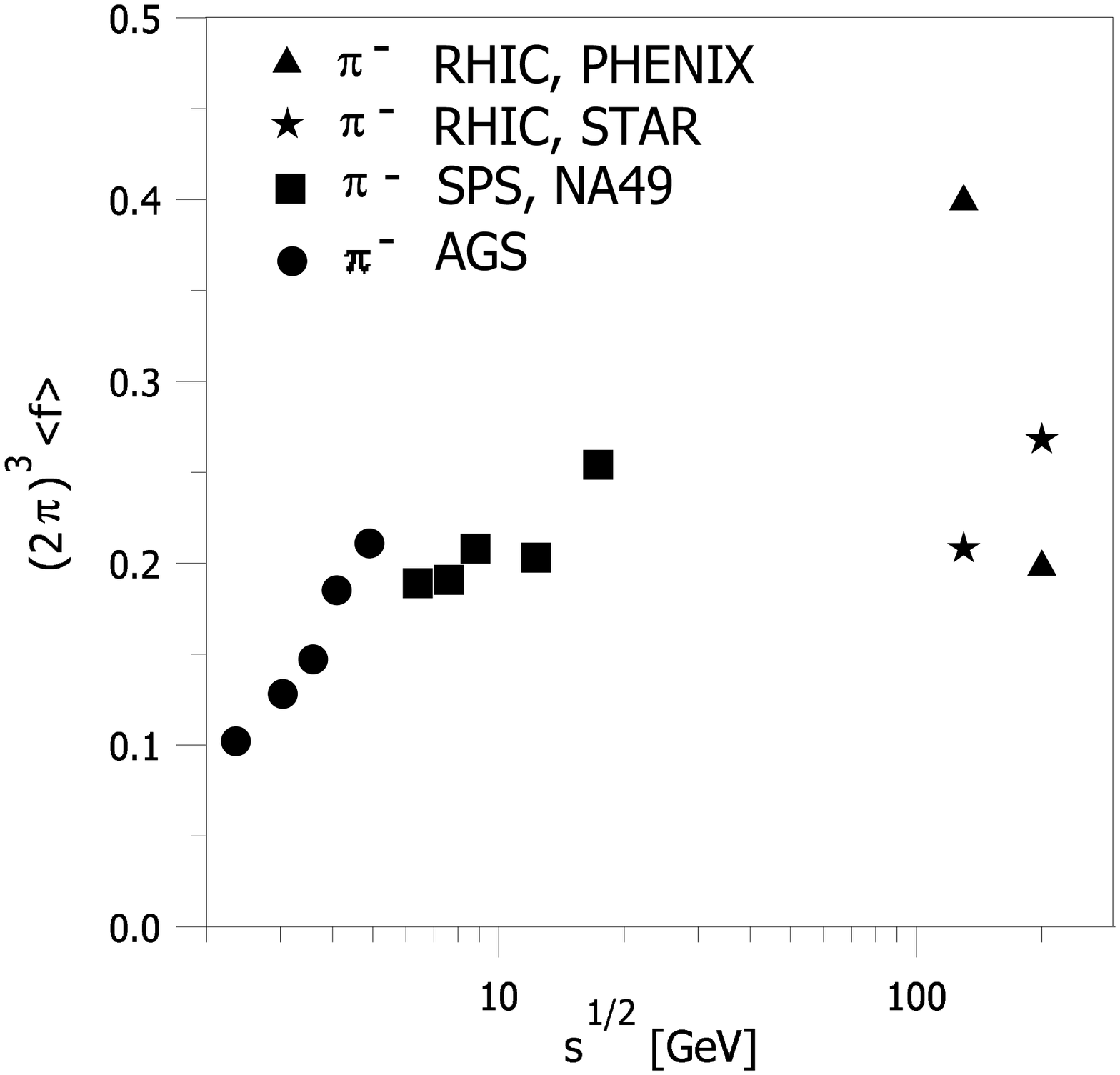}
\caption{ The average phase-space density of all negative  pions at
mid-rapidity,   $(2\pi )^{3}\left\langle f(y)\right\rangle$,
(circles, squares, stars and triangles) as function of c.m. energy
per nucleon in heavy ion central collisions.} \label{fig1}
\end{figure}

\newpage

\begin{figure}[h]
\centering
\includegraphics[scale=1.2]{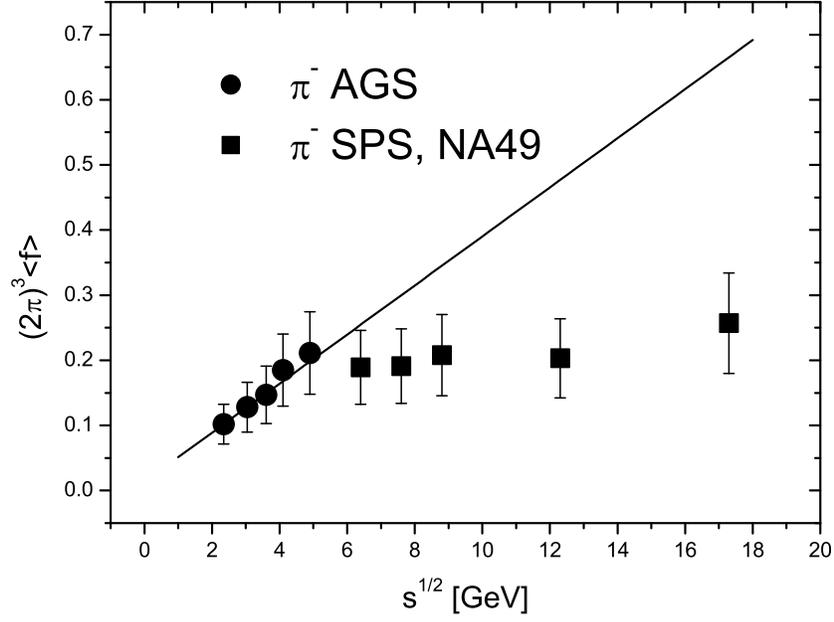}
\caption{The average phase-space density of all negative  pions at
mid-rapidity,   $(2\pi )^{3}\left\langle f(y)\right\rangle$,
 as function of c.m.
energy per nucleon at the AGS-SPS energy domain. The error bars take
into account uncertainties of the calculated APSD, see the text for
details. } \label{fig1}
\end{figure}

\newpage

\begin{figure}[h]
\centering
\includegraphics[scale=0.5]{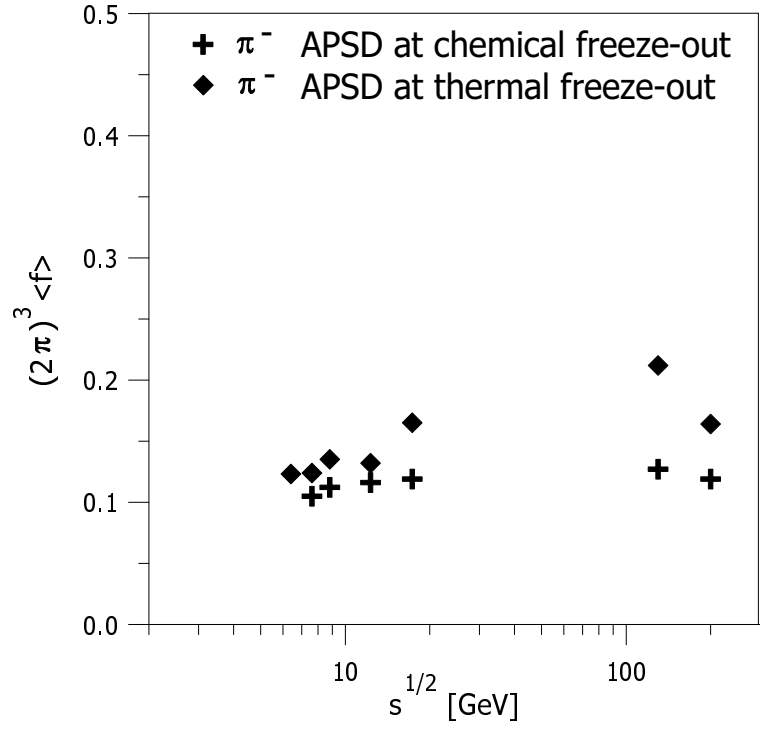}
\caption{ The average phase-space density of  thermal ("direct")
negative  pions, $(2\pi )^{3}\left\langle f(y)\right\rangle^{th}$
(rhombus), and the average phase-space density of negative pions at
the stage of chemical freeze-out, $(2\pi )^{3}\left\langle
f(y)\right\rangle^{ch}$ (crosses), as functions of c.m. energy per
nucleon in heavy ion central collisions.} \label{fig2}
\end{figure}

\newpage

\begin{figure}[h]
\centering
\includegraphics[scale=0.5]{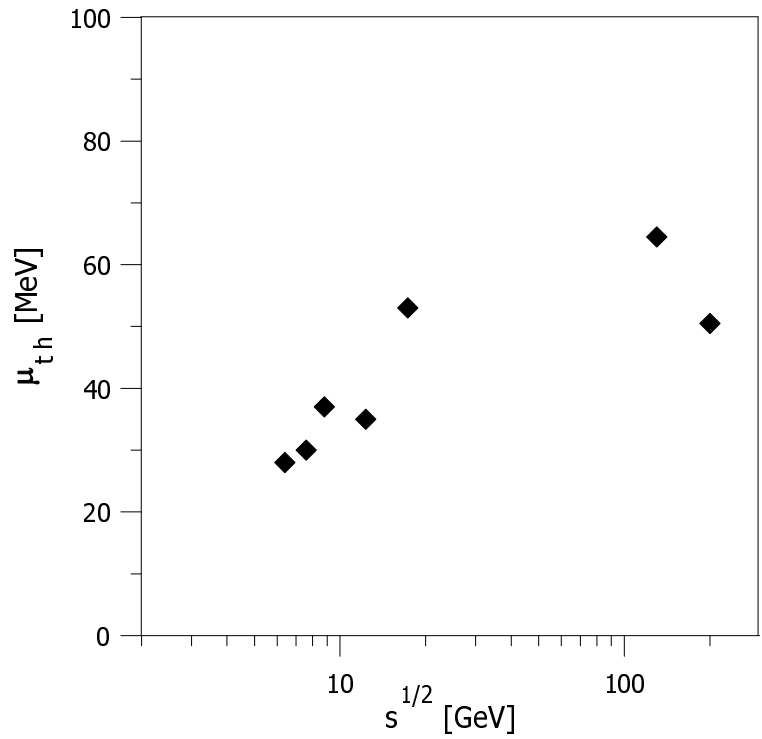}
\caption{ The chemical potential of  thermal ("direct") negative
pions, $\mu_{th}$, (rhombus) as function of c.m. energy per nucleon
in heavy ion central collisions.} \label{fig3}
\end{figure}

\newpage

\begin{figure}[h]
\centering
\includegraphics[scale=0.5]{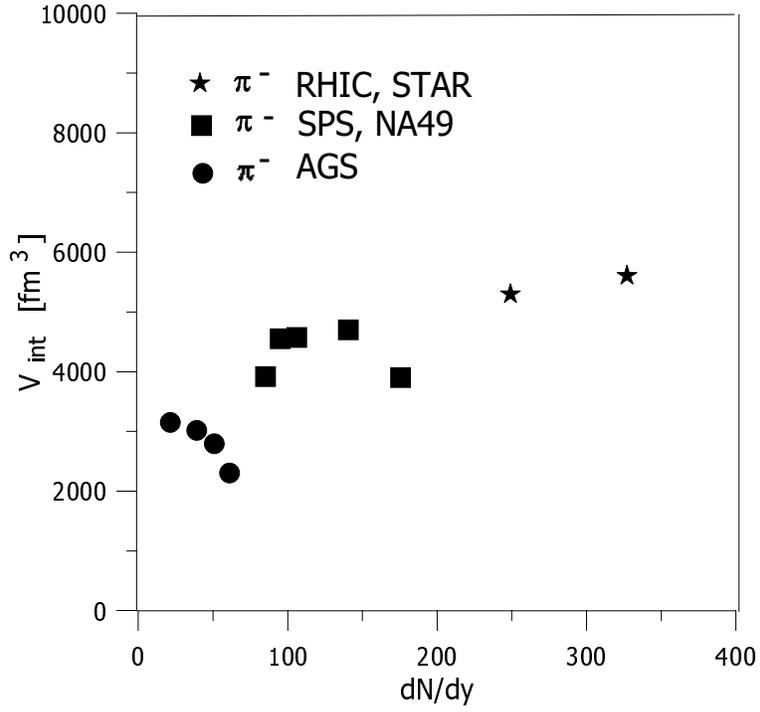}
\caption{ The interferometry volumes,
$V_{int}=(2\pi)^{3/2}R_{O}R_{S}R_{L}$, (circles, squares, and
stars) of negative pions at $p_{T}\simeq 0.06 - 0.07$ GeV  vs
rapidity densities  of the negative pions, $dN^{\pi^{-}}/dy$, at
mid-rapidity in  heavy ion central collisions at different
energies.} \label{fig4}
\end{figure}

\newpage

\begin{figure}[h]
\centering
\includegraphics[scale=0.5]{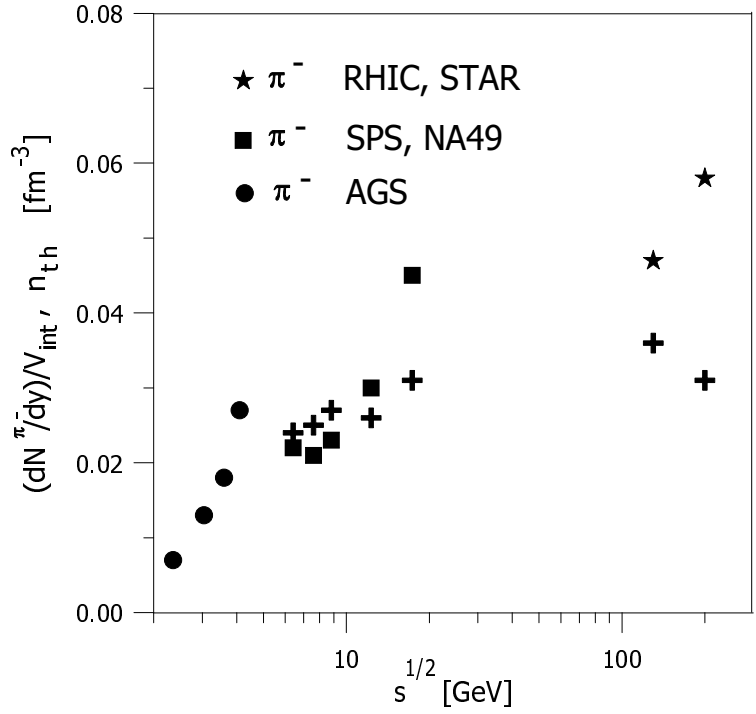}
\caption{ The ratio of  rapidity densities of all negative pions to
the corresponding interferometry volumes,
$(dN^{\pi^{-}}/dy)/V_{int}$,
 (circles, squares and  stars) and the ratio
of rapidity densities of negative thermal pions to their effective
volumes, that is thermal densities $n_{th}$, (crosses) vs c.m.
energies per nucleon in heavy ion central collisions.} \label{fig5}
\end{figure}

\newpage

\begin{figure}[h]
\centering
\includegraphics[scale=0.5]{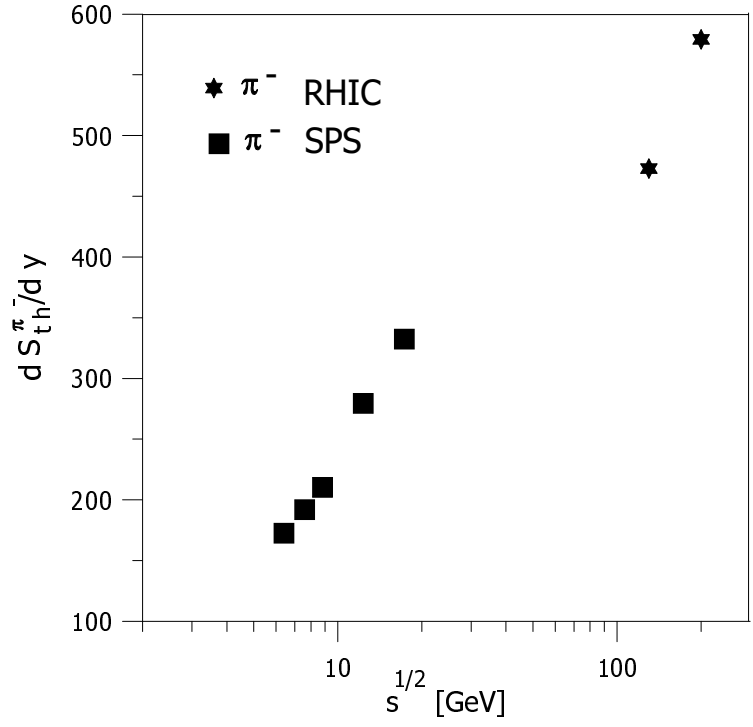}
\caption{ The  rapidity density of entropy for negative thermal
pions, $dS^{\pi^{-}}_{th}/dy$, (squares  and stars) as function of
c.m. energy per nucleon in heavy ion central collisions.}
\label{fig6}
\end{figure}

\newpage

\begin{figure}[h]
\centering
\includegraphics[scale=0.5]{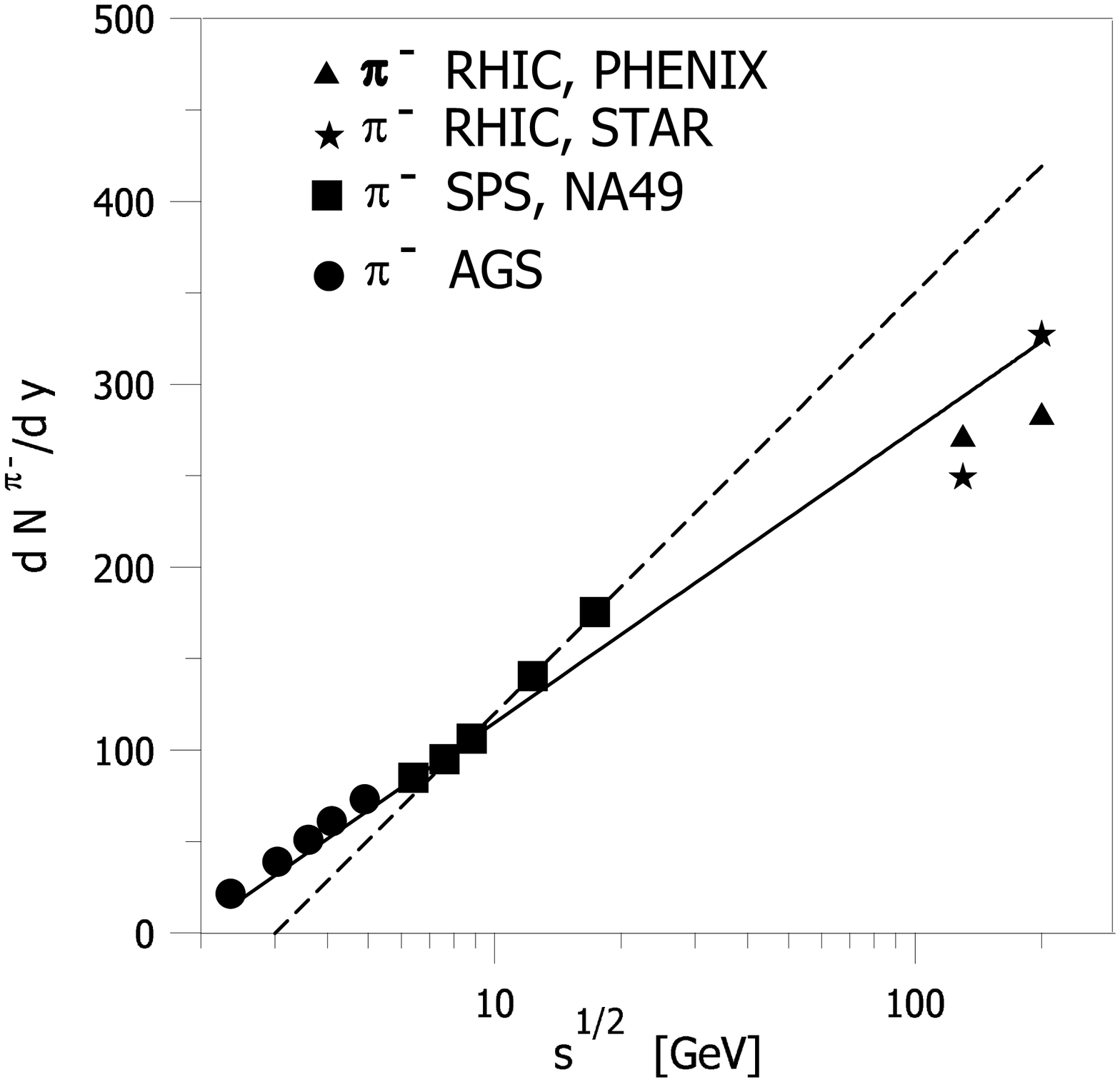}
\caption{ The  rapidity density of  negative pions,
$dN^{\pi^{-}}/dy$, as function of c.m. energy per nucleon in heavy
ion central collisions.} \label{fig7}
\end{figure}

\end{document}